\documentclass{iopart}

\usepackage{epsfig}
\newcommand{\beq}{\begin{equation}}
\newcommand{\eeq}{\end{equation}}
\newcommand{\beqn}{\begin{eqnarray}}
\newcommand{\eeqn}{\end{eqnarray}}
 
\newcommand\la{\langle}
\newcommand\ra{\rangle}
\newcommand\eps\varepsilon

\def\fm{\,\mbox{fm}}
\def\GeV{\,\mbox{GeV}}

\def\lsim{\mathrel{\rlap{\lower4pt\hbox{\hskip1pt$\sim$}}
    \raise1pt\hbox{$<$}}}         
\def\gsim{\mathrel{\rlap{\lower4pt\hbox{\hskip1pt$\sim$}}
    \raise1pt\hbox{$>$}}}    

\begin{document}

\begin{flushright}
{\small SLAC-PUB-10379\\
March, 2004}
\end{flushright}
\vspace*{-0.5cm}

\title{Heavy quark production and gluon shadowing at RHIC and LHC}
\author{J.~Raufeisen 
}
\address{Stanford Linear Accelerator Center, Stanford University,\\ 2575 Sand Hill Road, Menlo Park, CA 94025, USA}
\ead{jorg@slac.stanford.edu}

\begin{abstract}
I review the color dipole formulation of heavy quark production in the light of recent RHIC data. Since charm and bottom production directly probe the gluon density, these processes allow one to study shadowing and parton saturation at RHIC and LHC. The dipole approach provides a convenient framework to calculate these nuclear effects. I present numerical results for open charm and bottom production in proton-proton and proton-nucleus collisions and discuss transverse momentum broadening of heavy quarkonia.

\end{abstract}

\begin{figure}[b]
  \centerline{\scalebox{0.44}{\includegraphics{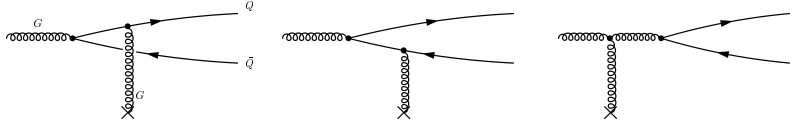}}}
    \center{
    {\caption{\em
      \label{fig:3graphs}The three lowest order
graphs contributing to heavy quark
 production in the dipole approach.  
These graphs correspond to the gluon-gluon
fusion mechanism of heavy quark production in the parton model.}  
    }  }
\end{figure}

At high center of mass energies 
$\sqrt{s}$, 
the cross section for any reaction $a+N\to\{b,c,\dots\}X$ can be expressed
as convolution of the light-cone (LC) wavefunction for the transition $a\to\{b,c,\dots\}$
and the cross section for scattering the color neutral
$\{{\rm anti-}a,b,c\dots\}$-system on the target nucleon $N$. In the case of heavy quark production this means that the Feynman graphs in Fig.~\ref{fig:3graphs} can be written in the form \cite{npz,kt},
\beq\label{eq:all}
\sigma(GN\to \{Q\bar Q\} X)
=\int_0^1 d\alpha \int d^2\rho 
\left|\Psi_{G\to Q\bar Q}(\alpha,\rho)\right|^2
\sigma_{q\bar q G}(\alpha,\rho),
\eeq 
where $\sigma_{q\bar qG}$ is
the cross section for scattering a color neutral quark-antiquark-gluon
system on a nucleon \cite{kt},
\beq\label{eq:qqG}
\sigma_{q\bar qG}(\alpha,\rho)
=\frac{9}{8}\left[\sigma_{q\bar q}(\alpha\rho)
+\sigma_{q\bar q}({\bar\alpha}\rho)\right]
-\frac{1}{8}\sigma_{q\bar q}(\rho).
\eeq
Here $\alpha$ is the light-cone momentum fraction carried by the 
heavy quark $Q$,
and $\bar\alpha=1-\alpha$
is the momentum fraction of the $\bar Q$. 
The flavor independent dipole cross section 
$\sigma_{q\bar q}(\rho)$ is a nonperturbative quantity and has to be determined from experimental data. It depends on the 
transverse separation $\rho$ between quark and antiquark. Note that the dipole approach is formulated in the target rest frame with the longitudinal axis parallel to the projectile gluon.
Higher order corrections cause $\sigma_{q\bar q}$ to depend on the target gluon momentum fraction $x_2$ as well, but this is not explicitly written out here. The LC wavefunctions $\Psi_{G\to Q\bar Q}$ can be calculated perturbatively, explicit expressions can be found {\em e.g.} in Ref.~\cite{kt}.

In momentum space, Eq.~(\ref{eq:all}) can be written in $k_\perp$-factorized form. Hence the dipole formulation is closely related to the approach of Ref.~\cite{Kirill}. However, the dipole approach takes into account only the finite transverse momentum of the target gluon. The relation of the dipole approach to the conventional parton model is explained in Ref.~\cite{Peng}.

\begin{figure}[t]
  \centerline{\scalebox{0.39}{\includegraphics{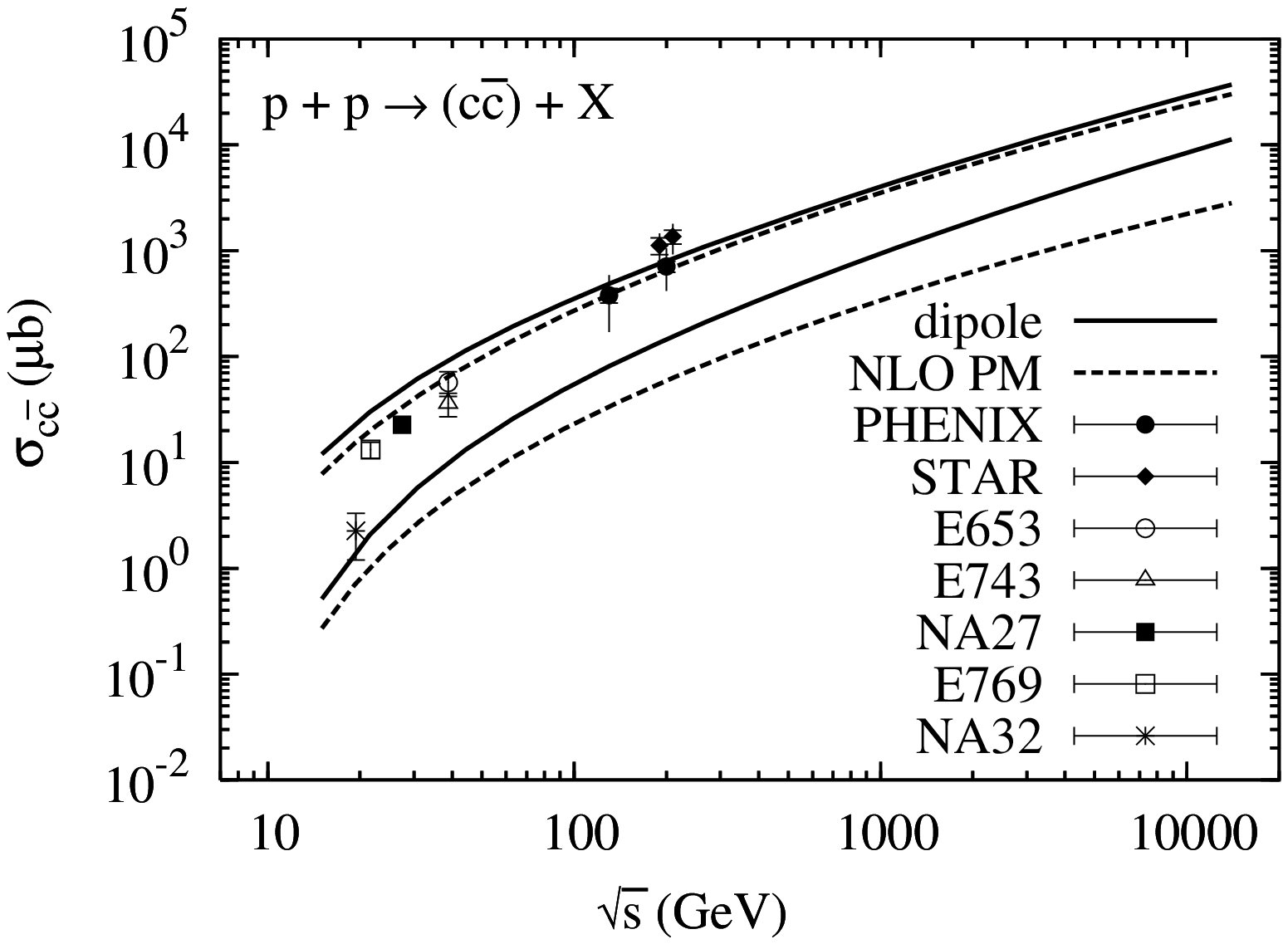}}
	      \scalebox{0.39}{\includegraphics{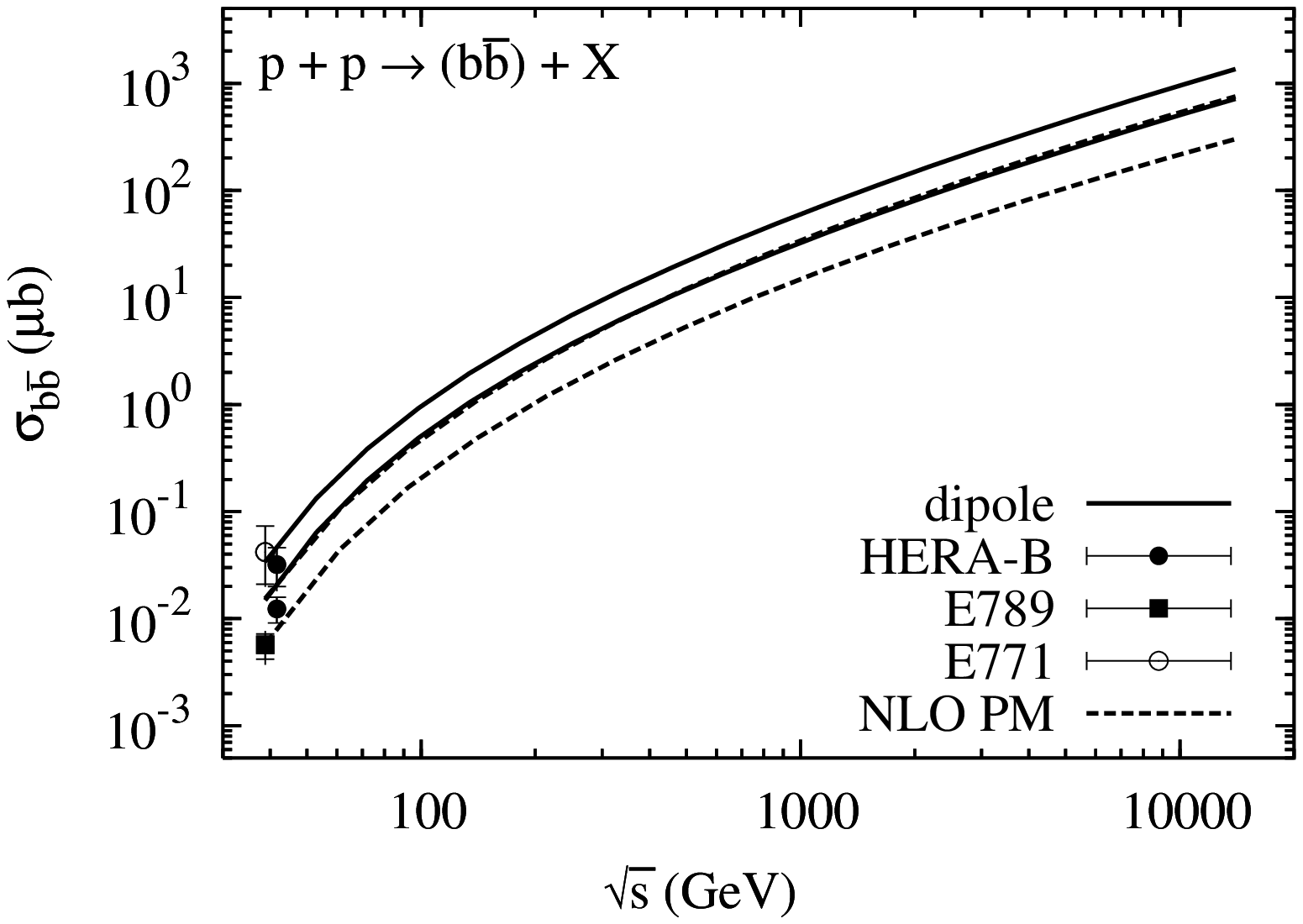}}}
\center{\parbox[thb]{13cm}{\caption{\em
      \label{fig:uncert}
Results for the total heavy quark pair cross section as function 
of cm.\ energy. Varying free parameters in the dipole approach (solid lines)
and in the NLO
parton model (dashed lines) \cite{nlo}
gives rise to the uncertainties. While the curves have already been published in Ref.~\cite{Peng}, this figure now also includes new preliminary data points \cite{data} from PHENIX and STAR (left panel, $\sqrt{s}=200\GeV$) and a
preliminary HERA-B measurement of open $b\bar b$ production (the lower HERA-B point.) }  
    }  }
\end{figure}

In order to calculate the cross section for heavy quark pair production
in proton-proton ($pp$) collisions,
Eq.~(\ref{eq:all}) has to be weighted with the projectile gluon density. Numerical results are shown in Fig.~\ref{fig:uncert}. Varying free parameters, such as the heavy quark mass $m_Q$, leads to the uncertainty represented by the space in between the solid curves (or in between the dashed curves in case of the next to leading order (NLO) parton model, respectively), see Ref.~\cite{Peng} for a complete discussion. 
As already pointed out in Ref.~\cite{Peng}, HERA-B energy is apparently too low for the dipole approach.

Partonic configurations with fixed transverse separations in
impact parameter space are known as eigenstates of the interaction. 
Therefore, the dipole approach is especially suitable to
describe multiple scattering effects, which become important at low $x_2$, when the lifetime of the $Q\bar Q$-pair becomes large enough to allow for more than one interaction with the target nucleus. There are two different sources of nuclear suppression for heavy quark production \cite{kt}:

First, the $Q\bar Q$ pair in Fig.~\ref{fig:3graphs} can rescatter several times inside the target. Since the amplitude for interaction with a single nucleon is known and because dipoles are interaction eigenstates, all these rescatterings can be resummed by eikonalizing $\sigma_{q\bar q}$ in Eq.~(\ref{eq:qqG}). Note that since the typical size of the pair is $\rho^2\sim 1/m_Q^2$, double scattering behaves parametrically like $\Delta\sigma/\sigma\propto Q_s^2(x_2)/m_Q^2$. Though formally higher twist, nuclear suppression due to heavy quark rescattering is enhanced by a factor of the saturation scale in the target nucleus, $Q_s^2(x_2)\propto A^{1/3}$ and cannot be neglected for charm quarks. 

In addition, there is the leading twist gluon shadowing \cite{kst2}. One of the quarks (or the gluons) in Fig.~\ref{fig:3graphs} can radiate another gluon, which then propagates together with the $Q\bar Q$-pair through the nucleus. The magnitude of the rescattering correction is determined by the transverse distance $\rho_G$ this gluon can propagate from its parent quark. Unlike the size of the heavy quark pair, $\rho_G$ is limited only by nonperturbative QCD effects. In Ref.~\cite{kst2}, these nonperturbative effects were modeled for the case of DIS by an attractive interaction between the gluon and the quark it is radiated off. In order to explain the smallness of the Pomeron-proton cross section, a value $\rho_G\lsim 0.3\fm$ is needed, resulting in a rather weak gluon shadowing in DIS. However, the concept of a nonperturbative interaction suggests that $\rho_G$ may depend on the color state the $Q\bar Q$-pair is in after radiation of the gluon \cite{kth,kt}. A pair in an octet state will interact more strongly with the gluon (small $\rho_G$), while the interaction would be much weaker between a color singlet pair and the gluon (large $\rho_G$). That way, gluon shadowing would be much stronger for charmonium production in $pA$ than in DIS. Here, I shall take a different point of view by arguing that $\rho_G$ is not determined by a nonperturbative interaction, but by  properties of the QCD vacuum that prevent the gluon from propagating distances $\gsim 0.3\fm$ (independent of the color state of the heavy quark pair). Gluon shadowing is then process independent, and is calculated as explained in the appendix of Ref.~\cite{krtj}. This viewpoint is supported by the observation of a rather weak suppression of the $J/\Psi$ yield in $dAu$ collisions \cite{Raphael}. However, these data do not disprove the predictions of Ref.~\cite{kth}, which are for $\chi$ mesons rather than for $J/\Psi$. 

Gluon shadowing $R_G(x_2,b)$ at impact parameter $b$ is included in the calculation through the modified eikonal approximation,
\beq\label{eq:eikonal}
\sigma_{q\bar q}^A(x_2,\rho)=2\int d^2b\left\{1-\exp\left(-\frac{1}{2}\sigma_{q\bar q}(x_2,\rho)T(b)R_G(x_2,b)\right)\right\},
\eeq
where $T(b)$ is the nuclear thickness. Note that the leading twist gluon shadowing has the effect of making the target nucleus more dilute, thereby reducing the saturation scale $Q_s^2$ and diminishing higher twist saturation effects.

\begin{figure}[t]
\centerline{\scalebox{0.38}{\includegraphics{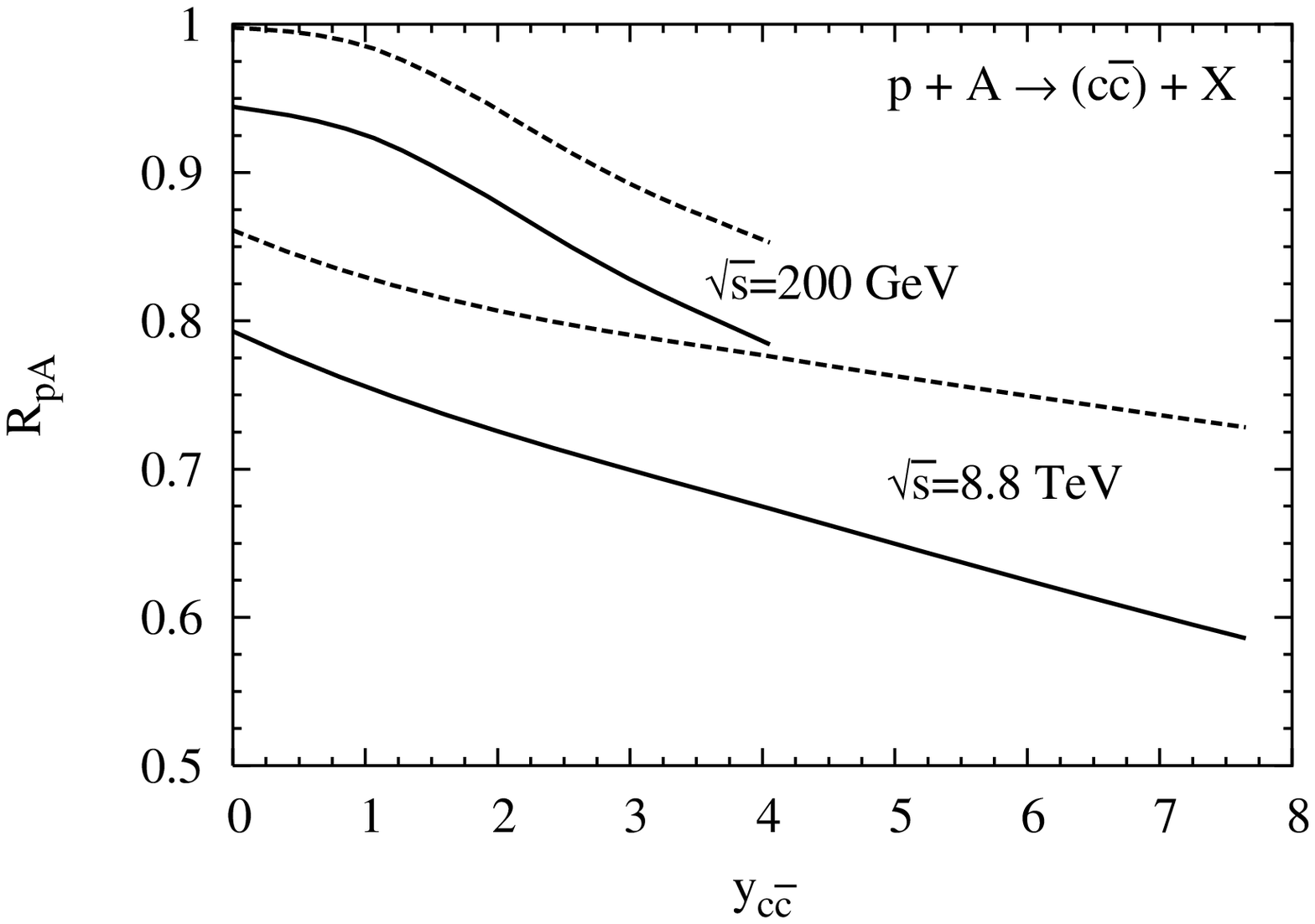}}
	      \scalebox{0.38}{\includegraphics{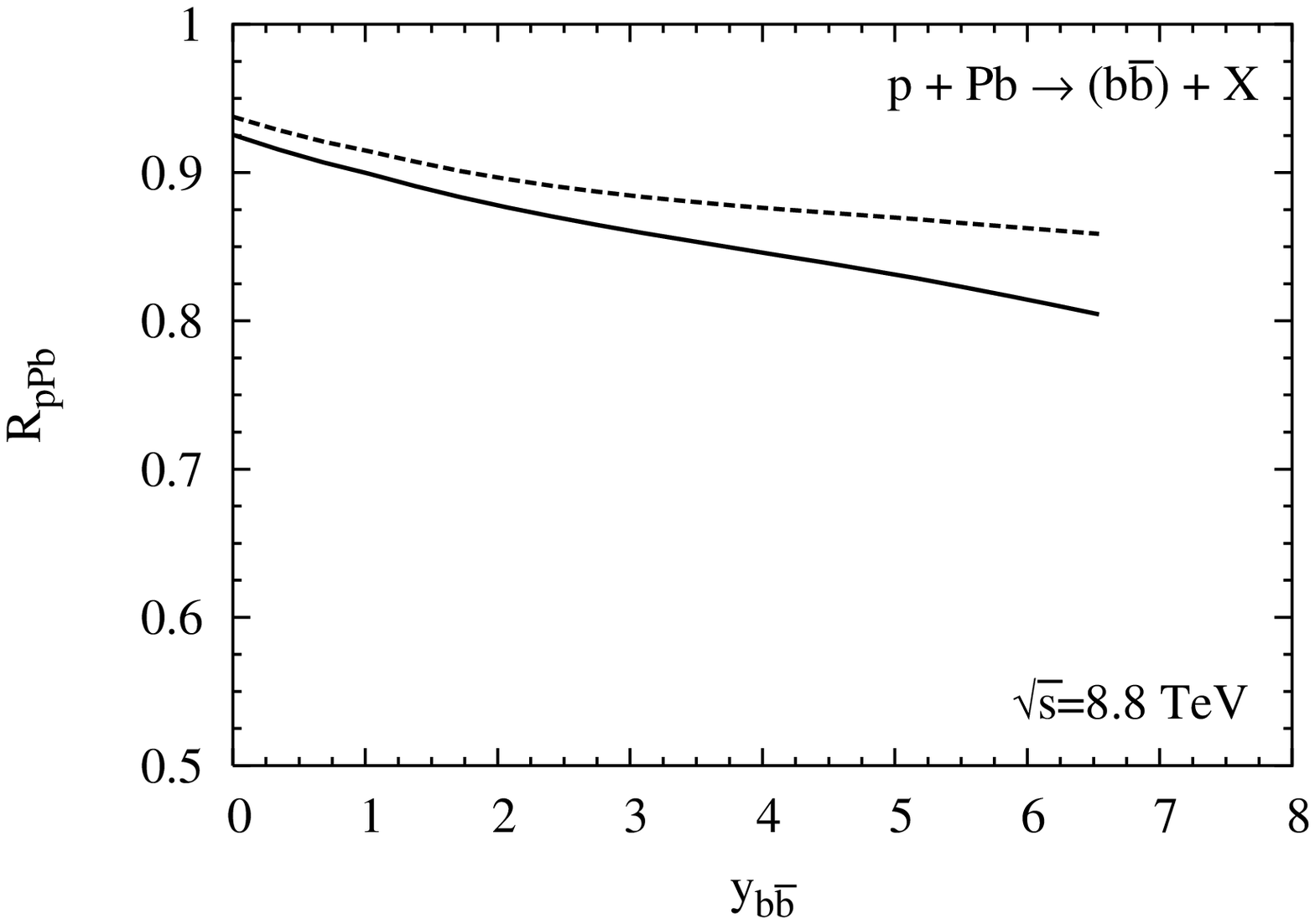}}}
    \center{
    {\caption{\em
      \label{fig:charm} Rapidity dependence of nuclear suppression for open heavy flavor production in $pAu$ ($\sqrt{s}=200 \GeV$) and in $pPb$ ($\sqrt{s}=8.8$ TeV) collisions. Dashed curves show gluon shadowing only, while solid curves also include rescattering of the heavy quark pair.}  
    }  }
\end{figure}

Results are shown in Fig.~\ref{fig:charm}. The eikonal approximation assumes that the lifetime of the heavy quark pair (the coherence length $l_c\propto1/x_2$) is much larger than the nuclear radius $R_A$. This is certainly the case at the LHC ($\sqrt{s}=8.8$ TeV). For $y_{c\bar c}=0$ at RHIC, however, $l_c\sim R_A$ and a more sophisticated calculation using the Green's function technique developed in Ref.~\cite{green} would yield a slightly smaller suppression. 
Gluon shadowing sets in at smaller $x_2$ than quark shadowing, since the $Q\bar QG$-fluctuation lives much shorter than the $Q\bar Q$-pair alone (because of the larger invariant mass of the $Q\bar QG$-state \cite{lc}.) This is taken into account in this calculation. In particular there is no gluon shadowing at $y_{c\bar c}=0$ at RHIC. The rescattering of the heavy quark pair is a sizable contribution to the nuclear suppression of open charm production, especially at $\sqrt{s}=200\GeV$, where gluon shadowing is weak. This is one reason why the observed suppression in $J/\Psi$ yields in $dAu$ collisions at RHIC cannot be explained by a modification of the nuclear gluon distribution alone. The rescattering of $b\bar b$-pairs at the LHC is expected to be only a small effect.

The PHENIX collaboration also reported values for transverse momentum broadening of $J/\Psi$ mesons in $dAu$ collisions \cite{Raphael} that are significantly larger than in fixed target experiments. An increase of $\delta\la p_T^2\ra$ with energy is expected in the dipole approach \cite{broad,broadletter} but the relevant energy scale is $\sqrt{x_1s}$ rather than the hadronic $cm.$ energy $\sqrt{s}$. For RHIC kinematics, $\sqrt{x_1s}$ is even smaller than in $800\GeV$ fixed target experiments. The predictions from Ref.~\cite{broadletter}, assuming incoherent initial state rescattering of the projectile gluon, are $\delta\la p_T^2\ra(x_2=0.09)=0.4\GeV^2$ (exp: $(1.77\pm0.35)\GeV^2$) and
$\delta\la p_T^2\ra(x_2=0.003)=0.8\GeV^2$ (exp: $(1.29\pm0.35)\GeV^2$). It is unclear what causes the large discrepancy at $x_2=0.09$. At smaller $x_2$, there could be an additional contribution to $\delta\la p_T^2\ra$ from the color filtering mechanism of Ref.~\cite{krtj}.

\ack{I am grateful to Mikkel Johnson, Boris Kopeliovich and Jen-Chieh Peng for valuable discussion. This research was supported in part under the Feodor Lynen Program of the Alexander von Humboldt Foundation and by the U.S.\ Department of Energy under Contract No.\ DE-AC03-76SF00515.
}


\section*{References}
\begin{thebibliography}{20}

\bibitem{npz}
N.~N.~Nikolaev, G.~Piller and B.~G.~Zakharov,
J.\ Exp.\ Theor.\ Phys.\  {\bf 81}, 851 (1995)
[Zh.\ Eksp.\ Teor.\ Fiz.\  {\bf 108}, 1554 (1995)];
Z.\ Phys.\ A {\bf 354}, 99 (1996).

\bibitem{kt}
B.~Z.~Kopeliovich and A.~V.~Tarasov,
Nucl.\ Phys.\ A {\bf 710}, 180 (2002).

\bibitem{Kirill}
K.~Tuchin,
arXiv:hep-ph/0401022;
D.~Kharzeev and K.~Tuchin,
arXiv:hep-ph/0310358;
F.~Gelis and R.~Venugopalan,
Phys.\ Rev.\ D {\bf 69}, 014019 (2004).

\bibitem{Peng}
J.~Raufeisen and J.~C.~Peng,
Phys.\ Rev.\ D {\bf 67}, 054008 (2003).

\bibitem{nlo}
P.~Nason, S.~Dawson and R.~K.~Ellis,
Nucl.\ Phys.\ B {\bf 303}, 607 (1988);
Nucl.\ Phys.\ B {\bf 327}, 49 (1989);
M.~L.~Mangano, P.~Nason and G.~Ridolfi,
Nucl.\ Phys.\ B {\bf 373}, 295 (1992).

\bibitem{data}
S.~Kelly (PHENIX), A.~Tai (STAR) and J.~Spengler (HERA B), talks presented at Quark Matter 2004 conference, Oakland, CA, 11-17 Jan.\ 2004.
 
\bibitem{kst2}
B.~Z.~Kopeliovich, A.~Sch\"afer and A.~V.~Tarasov,
Phys.\ Rev.\ D {\bf 62}, 054022 (2000).

\bibitem{kth}
B.~Z.~Kopeliovich, A.~V.~Tarasov and J.~H\"ufner,
Nucl.\ Phys.\ A {\bf 696}, 669 (2001).

\bibitem{krtj}
B.~Z.~Kopeliovich, J.~Raufeisen, A.~V.~Tarasov and M.~B.~Johnson,
Phys.\ Rev.\ C {\bf 67}, 014903 (2003).

\bibitem{Raphael}
M.~Brooks and R.~de~Cassagnac (PHENIX), talks  presented at Quark Matter 2004 conference, Oakland, CA, 11-17 Jan.\ 2004.

\bibitem{green}
B.~Z.~Kopeliovich, J.~Raufeisen and A.~V.~Tarasov,
Phys.\ Lett.\ B {\bf 440}, 151 (1998);
J.~Raufeisen, A.~V.~Tarasov and O.~O.~Voskresenskaya,
Eur.\ Phys.\ J.\ A {\bf 5}, 173 (1999).

\bibitem{lc}
B.~Z.~Kopeliovich, J.~Raufeisen and A.~V.~Tarasov,
Phys.\ Rev.\ C {\bf 62}, 035204 (2000).

\bibitem{broad}
M.~B.~Johnson, B.~Z.~Kopeliovich and A.~V.~Tarasov,
Phys.\ Rev.\ C {\bf 63}, 035203 (2001).

\bibitem{broadletter}
J.~Raufeisen,
Phys.\ Lett.\ B {\bf 557}, 184 (2003).

\end{thebibliography}
\end{document}